\documentclass[reprint,pra,showpacs,preprintnumbers,amsmath,amssymb]{revtex4-1}
\usepackage{graphicx}
\usepackage{dcolumn}
\usepackage{amsmath}
\usepackage{times}
\usepackage{mathtools}
\usepackage{rotating}

\begin{document}

\title{Relativistic Spin-$0$ Feshbach-Villars Equations for Polynomial Potentials }

\author{B.~M.~Motamedi}
\author{T.~N.~Shannon}
\author{Z.~Papp}
\affiliation{ Department of Physics and Astronomy,
California State University Long Beach, Long Beach, California, USA }

\date{\today}

\begin{abstract}

We propose a solution method for studying relativistic spin-$0$ particles. We adopt the Feshbach-Villars formalism of the 
Klein-Gordon equation and express the formalism in an integral equation form. 
The integral equation is represented in the Coulomb-Sturmian basis. 
The corresponding Green's operator with  Coulomb and linear confinement potential can be calculated as a matrix continued fraction.
We consider Coulomb plus short range vector potential for bound and resonant states and linear confining scalar potentials
for bound states.
The continued fraction is naturally divergent at resonant state energies, but we made it convergent by an appropriate analytic continuation. 

\end{abstract}


\maketitle

\section{Introduction}
\label{intro}

The Klein-Gordon equation is  the basic relativistic quantum mechanical equation of spin-$0$ particles. 
However, it contradicts some of the postulates of quantum mechanics. 
In quantum mechanics, it is postulated that the system is 
completely determined by the wave function and the time evolution of the wave function is determined 
by the time-dependent  Schr\"odinger equation. The Klein-Gordon
equation is second order in time derivative. Therefore, to determine the system uniquely  we need its time derivative as well. 

In order to give a proper interpretation  Feshbach and Villars  rewrote the Klein-Gordon equation in  Hamiltonian form
 \cite{Feshbach:1958wv}. In  the Feshbach-Villars (FV0)  formalism we split
the Klein-Gordon wave function into two components, and  for the components vector we
arrive at a Schr\"odinger-like equation with a first order time derivative.  The equations look like usual coupled differential
equations, but the components are coupled by the kinetic energy operator, which makes them difficult to solve. A few demonstrations of 
their use are in 
Refs.\ \cite{fuda1980feshbach,friar1980feshbach,fuda1981inversion,merad2000boundary,khounfais2004scattering,bounames2001solution,horbatsch1995analysis}.

In a recent work we have proposed a solution method for the Feshbach-Villars equations \cite{brown2016matrix}. 
The method amounts to rewriting
the eigenvalue problem 
into an integral equation form and representing the equation on a discrete Hilbert space basis. The Green's operator has
been calculated by a matrix continued fraction. However, the continued fraction in Ref.\ \cite{brown2016matrix} converges only 
for bound state energies.

The aim of this work is to further develop the solution method of
spin-zero  FV0 equations to include  resonant states and  confining potentials.
In Sec.\ II, we outline the FV0 formalism. Then, in Sec.\ III we 
recapitulate the solution method of  Ref.\ \cite{brown2016matrix}. In Sec.\ IV we perform the analytic continuation and
in Sec.\ V we examine the case of linear confining potential.
Finally we summarize our findings and draw some conclusions. \\

\section{Feshbach-Villars equations for spin-zero particles}
\label{fv0-outline}

The Klein-Gordon equation for a free spin-$0$ particle is given by
\begin{equation}
-\hbar^{2} \frac{\partial^{2}}{\partial t^{2}}    \Psi  = \left( c^{2} p^{2} + m^{2}c^{4} \right)  \Psi .
\end{equation}
We can introduce interaction by minimal coupling  $p_{\mu} \to p_{\mu}  -  q/c\,  A_{\mu}$, where $p_{\mu}$ and $A_{\mu}$ are the 
four-momentum and the four-potential, respectively. This interaction transforms like a four-vector with respect to the
Lorentz transformation. 
We can also introduce a scalar interaction $S$ by the substitution $m \to m + S/c^{2}$, 
which is basically a position dependent effective mass.
 So, if we take $\vec{A}=0$ and denote the fourth component of the vector potential by $V$, we have
\begin{equation}
\left(i\hbar  {\partial}/{\partial t}  - V \right)^{2}  \Psi  = \left[ c^{2} p^{2} + (m + S/c^{2})^{2}c^{4} \right]  \Psi .
\end{equation}
In the FV0 formalism the wave function is split into two components
\begin{eqnarray}
\Psi &=&  \phi + \chi,  \\
\left( i \hbar \frac{\partial }{ \partial t} - V \right) \Psi & = & m c^{2} (\phi -\chi),
\end{eqnarray}
and for the components we can readily derive the coupled equations
\begin{eqnarray}
i\hbar \frac{\partial}{\partial t} \phi & = & \left(  \frac{p^{2}}{2m} +U \right)  (\phi + \chi) + ( mc^{2} +V) \phi ~,  \label{fv1} \\
i\hbar \frac{\partial}{\partial t} \chi & = & - \left(  \frac{p^{2}}{2m} +  U \right) (\phi + \chi) - ( mc^{2} -V) \chi   ~,  \label{fv2}  
\end{eqnarray}
where $U = S + {S^{2}}/{2mc^{2}}$.

If we introduce the two-component wave function
\begin{equation}
| \psi \rangle = \begin{pmatrix}  \phi \\ \chi \end{pmatrix} ,
\end{equation}
we can define the Hamiltonian
\begin{equation} \label{HFV0}
H_{FV0}  =  (\tau_{3}+i\tau_{2}) \frac{p^{2}}{2m} +\tau_{3} mc^{2} + (\tau_{3}+i\tau_{2}) U +  V ,
\end{equation}
where $\tau_{i}$  denote the Pauli matrices 
\begin{equation}
\tau_{1}= \begin{pmatrix}  0 & 1 \\ 1 & 0 \end{pmatrix}, \ \ \ \tau_{2}= \begin{pmatrix}  0 & -i \\ i & 0 \end{pmatrix}, \ \ \
\tau_{3}= \begin{pmatrix}  1 & 0 \\ 0 & -1 \end{pmatrix}.
\end{equation}
Now we can write (\ref{fv1}) and (\ref{fv2}) into a form analogous to the time-dependent Schr\"odinger equation
\begin{equation}
i \hbar \frac{\partial}{\partial t} |\psi \rangle = H_{FV0} | \psi \rangle,
\end{equation}
or, for stationary states we have
\begin{equation}
  H_{FV0} | \psi \rangle = E | \psi \rangle ~.
  \label{hpsi}
\end{equation}

The Hamiltonian $H_{FV0}$ of Eq.\ (\ref{HFV0}) is
Hermitian in the generalized sense 
\begin{equation}
H_{FV0} = \tau_{3} H_{FV0}^{\dagger} \tau_{3},
\end{equation}
and it has real eigenvalues  \cite{Feshbach:1958wv,wachter2010relativistic}. The wave function is normalized according to
\begin{equation}
\langle \psi | \tau_{3} | \psi \rangle = \pm 1,
\end{equation}
where the plus or minus sign corresponds to particle or antiparticle.

This Hamiltonian looks like a usual coupled-channel Hamiltonian. 
However, in a usual coupled-channel Hamiltonian
the channels are coupled by a short-range potential. Here the coupling is due to the kinetic energy operator, which is not
a short range operator and cannot be neglected even at asymptotic distances.   
 This may be the reason why the FV0 equations are not used frequently in practical calculations.  
 If we want to solve the FV0 equations in a proper way we should not approximate the kinetic energy operator 
 even in the coupling terms.

\section{Solution method  }
\label{nonrel-outline}

In order to solve the equations we need to write the Hamiltonian as
\begin{equation}
H_{FV0} =H_{FV0}^{(l)}  + H_{FV0}^{(s)},
\end{equation}
where $H_{FV0}^{(l)}$ is the asymptotically relevant long range, and  $H_{FV0}^{(s)}$ is the asymptotically irrelevant
short range part. If we make a similar separation of the potentials,
\begin{equation}
V = V^{(l)}+ V^{(s)} \quad \text{and} \quad U = U^{(l)}+ U^{(s)},
\end{equation}
we can write
\begin{equation}
H_{FV0}^{(l)} =  (\tau_{3}+i\tau_{2}) \frac{p^{2}}{2m} +\tau_{3} mc^{2} + (\tau_{3}+i\tau_{2}) U^{(l)} +  V^{(l)} 
\end{equation}
and
\begin{equation}
H_{FV0}^{(s)} =    (\tau_{3}+i\tau_{2}) U^{(s)} +  V^{(s)} .
\end{equation}
Then we can cast the eigenvalue problem of Eq.\ (\ref{hpsi}), for bound and resonant states,  into a Lippmann-Schwinger form
\begin{equation}
| \psi \rangle =  G_{FV0}^{(l)}(E) H_{FV0}^{(s)}   | \psi \rangle ,
\label{LSeq}
\end{equation}
where $ G_{FV0}^{(l)}(E) = ( E -  H_{FV0}^{(l)})^{-1}$ is the Green's operator of the long-range FV0 Hamiltonian.

A natural way of solving this integral equation is to approximate $H_{FV0}^{(s)}$ on a finite subset of a discrete basis.
This results in a finite-rank expansion of the short range term. Several  expansion schemes have been proposed. Recently we 
have found a simple, straightforward, yet very efficient approximation scheme  \cite{brown2013approximations}. 
We need to represent the short-range potential in a larger basis, invert the potential matrix, truncate to a smaller basis,
and then invert it back. This way we achieve a low-rank representation of the potential operator that  
contains the relevant information from the larger basis. With this approach, even a low-rank representation gives very good results,
while higher-rank representations provide extremely accurate results.
If we represent the short-range interaction by an $N\times N$ basis, then for solving Eq.\ (\ref{LSeq}) we need only an 
$N\times N$ representation of $G_{FV0}^{(l)}$.

The Green's operator $G$ satisfies the operator equation
\begin{equation}
J  G = G J = 1,
\end{equation}
where $J=(E-H)$. The evaluation of the Green's operator basically amounts to inverting an 
infinite matrix, which is, in general, rather complicated. 
A notable exception is when the basis is such that $J$ appears as an infinite tridiagonal matrix \cite{Konya:1997JMP,PRADemir2006}.
In this case
\begin{equation}
\underline{G} = \left( \underline{J} - \delta_{iN}\delta_{jN} J_{N,N+1} C_{N+1} J_{N+1,N} \right)^{-1},
\label{G1}
\end{equation}
where the underline denotes $N \times N$ matrices and $C$ is a continued fraction.
Basically, $\underline{G}$ is almost the inverse of $\underline{J}$, only the right-lower corner of
$\underline{J}$ is modified by a continued fraction. The continued fraction is
 constructed from the higher-index elements of $J$ and is 
defined by the recursion relation
\begin{equation}\label{cf}
C_{N+1}= (J_{N+1,N+1} - J_{N+1,N+2} C_{N+2}  J_{N+2,N+1})^{-1}.
\end{equation}

The validity of this approach has been established for infinite tridiagonal $J$ matrices  \cite{Konya:1997JMP}. 
The derivation is based on 
the intimate relation between three-term recursion relations and continued fractions. So, if $J$ is a band matrix, such as
penta-diagonal or septa-diagonal, the method is not applicable. However, all band matrices can be considered as tridiagonal 
matrices of block matrices. Therefore the above procedure is applicable \cite{kelbert2007green}. 
The only difference is that $J_{i,j}$ is not a number 
any more, but rather an $m \times m$ block matrix and   $C$ therefore is a matrix continued fraction. 

In matrix representation Eq.\ (\ref{LSeq}) becomes a homogeneous algebraic equation
\begin{equation}
\left[ \left(  \underline{G}_{FV0}^{(l)}(E) \right)^{-1} - \underline{H}_{FV0}^{(s)}  \right] \underline{ \psi}  =  0 ,
\label{LSeq1}
\end{equation}
which is solvable if the determinant of the expression in the bracket vanishes.

\subsection{The basis}

As a basis, we adopted the Coulomb-Sturmian (CS) functions
\begin{equation}
\langle r | n \rangle = \left(\frac{n!}{(n+2l+1)!} \right)^{1/2} e^{-br}(2br)^{l+1}L_{n}^{2l+1}(2br),
\end{equation}
where $l$ is the angular momentum, $L$ is the Laguerre polynomial, and $b$ is a parameter. 
They also have a nice form in momentum representation
\begin{equation}
\begin{split}
\langle p | n \rangle =  & \frac{2^{l+ 3/2}l!(n+l+1)\sqrt{n!}}{\sqrt{\pi(n+2l+1)!}}   \frac{b(2bp)^{l+1}} {(p^{2}+b^{2})^{2l+2}} \\
 & \times G_{n}^{l+1} \left(\frac{p^{2}-b^{2}}{p^{2}+b^{2}}\right),
\end{split}
\end{equation}
where $G$ is the Gegenbauer polynomial. 

The CS functions satisfy Sturm-Liouville type  differential equations
\begin{equation}
\left(  - \frac{d^{2}}{dr^{2}}  + \frac{l(l+1)}{r^{2}}  - \frac{2b(n+l+1)  }{r}  +b^{2}\right)  \langle r | n \rangle = 0.
\end{equation}
Consequently with $\langle r |  \widetilde{n} \rangle = \langle r | n \rangle /r$ 
we have the orthogonality $\langle \widetilde{n}  |n' \rangle = \delta_{n n'}$
and the completeness relation
\begin{equation}
1 = \lim_{N\to\infty } \sum_{n=0}^{N} | n \rangle \langle \widetilde{n} | = \lim_{N\to\infty } \sum_{n=0}^{N} | \widetilde{n} \rangle \langle n | ~.
\end{equation}

\subsection{Matrix elements }

The simple form of the CS basis allows the exact and analytic calculation of the matrix
elements
\begin{widetext}
\begin{equation}
\langle n  |  {1}/{r} | n' \rangle  = \delta_{n n'},
\label{n1/rn}
\end{equation}
\begin{equation}
 \langle n  |n' \rangle  = \langle n'  |n \rangle = 
	\begin{dcases}  (n+l+1)/b  &   n=n', \\
		-\sqrt{n'(n'+2l+1)}/(2b) & n'=n+1, \\
		0  & n' > n+ 1,
	\end{dcases}
\label{nn}
\end{equation}
\begin{equation}
 \langle n  |  {p^{2}}|n' \rangle =  \langle n'  | {p^{2}}|n \rangle  
	=\begin{dcases}  (n+l+1)\:b  &   n=n', \\
		\sqrt{n'(n'+2l+1)}\:b/2 & n'=n+1, \\
		0  &  n' > n+1,
	\end{dcases}
\label{np2n}
\end{equation}
\begin{equation}
  \langle {n } | r | {n'} \rangle  = \langle {n'} | r | {n} \rangle   
 = \begin{dcases} 
( 6n^{2} +2(l+1)(6n+2l+3  ) )/(4b^{2})  & n'=n,\\ 
- (2 n' + 2l+1  )  \sqrt{n'(n'+2l+1)} /(2b^{2}) & n' =n+1,\\
  \sqrt{n'(n'-1)(n'+2l)(n'+2l+1)} /4b^{2}& n'= n+2,\\
0 & n'> n+2,
\end{dcases} 
\label{nrn}
\end{equation}
and
\begin{equation}
 \langle {n } | r^{2} | {n'} \rangle  = \langle {n'} | r^{2} | {n} \rangle   
= \begin{dcases} 
   [(((10n+2l+4)(n+2l+3)+9n(n-1))(n+2l+2)+n(n-1)(n-2)) ]/(8 b^{3})  & n'=n, \\ 
- [ (4n'+2l)(n'+2l+2)+(n'-1)(n'-2) ]\sqrt{n'(n'+2l+1)} \,  3/(8b^{3}) & n' =n+1, \\
   (2n'+2l)\sqrt{n'(n'-1)(n'+2l+1)(n'+2l) } \, 3/(8b^{3} ) & n'= n+2, \\
   -\sqrt{n'(n'-1) (n'-2) (n'+2l+1)(n'+2l)(n'+2l-1 ) } /(8b^{3} ) & n'= n+3, \\
0 & n'> n+3.
\end{dcases} 
\label{nr2n}
\end{equation}
\end{widetext}

\section{Coulomb plus short range potential }

We  assume here that  the fourth component of the vector potential  is Coulomb-like 
\begin{equation}
V =  {Z}/{r} + v_{4}^{(s)}.
\end{equation}
Consequently we have
\begin{equation}
H_{FV0}^{(s)} =  v_{4}^{(s)},
\end{equation}
whose CS matrix elements can easily be determined.

The real difficulty lies in the evaluation of  the CS matrix representation of $G_{FV0}^{(l)}$.
From Eq.\ (\ref{HFV0}) we have
\begin{equation}
J = E  -  H_{FV0}^{(l)} = E -  \left( (\tau_{3}+i\tau_{2}) \frac{p^{2}}{2m} +\tau_{3} mc^{2}   +  \frac{Z}{r} \right).
\end{equation}
We have learned before that in the CS basis the constants $E$ and $mc^{2}$, the $p^{2}$ and the $1/r$ terms are either tridiagonal or diagonal 
$\infty \times \infty$ matrices. 
On the other hand, due to the matrix structure of the FV0 equations, a $2\times 2$ matrix structure becomes 
superimposed on the tridiagonal structure. Therefore, $J$ is a block tridiagonal matrix with $2\times 2$ blocks. 
As a result Eq.\ (\ref{G1}) is applicable and $C_{N+1}$ becomes a matrix continued fraction with $2\times 2$ blocks.

This method has been used in Ref.\ \cite{brown2016matrix} for calculating bound states. 
The matrix continued fraction was evaluated iteratively. Assuming that $C_{N'}=0$ for $N' >>N$, the continued fraction 
in Eq.\ (\ref{cf}) was evaluated 
backwards. However, this procedure converges only for bound state energies.

\subsection{Analytic continuation of the Coulomb Green's matrix}

In order to extend the method for resonant states we need to perform analytic continuation of the matrix continued fraction. 
For this purpose we need to estimate the tail of the continued fraction. 
We can see from Eqs.\ (\ref{n1/rn}-\ref{np2n}) that  $J_{N'+1,N'+1}   \simeq  J N'$
with
\begin{equation}  
J =  \begin{pmatrix}  \frac{E}{b} -  \frac{mc^{2}}{b} -  \frac{\hbar^{2}b}{2m} & -  \frac{\hbar^{2}b}{2 m}   \\ 
  \frac{\hbar^{2}b}{2m}  &   \frac{E}{ b} +   \frac{mc^{2}}{b} +  \frac {\hbar^{2}b}{2m}  \end{pmatrix} 
\end{equation}
and $J_{N' ,N'+1} \simeq J' N' $ with
\begin{equation}  
J' =  \begin{pmatrix}  - \frac{E}{2b} +  \frac{mc^{2}}{2b} -  \frac{\hbar^{2}b}{4m} & -  \frac{\hbar^{2}b}{4 m}   \\ 
  \frac{\hbar^{2}b}{4m}  &  - \frac{E}{2 b} -   \frac{mc^{2}}{2b} +   \frac{\hbar^{2}b}{4m}  \end{pmatrix} 
\end{equation}
as $N' \to \infty$. Then from Eq.\ (\ref{cf}) it also follows that $C_{N'+1} \simeq C / N'$. So, as $N'\to \infty$ we find
\begin{equation}
C = (J  - J' C  J')^{-1}.
\label{Ceq}
\end{equation} 
With a little manipulation we obtain
\begin{equation}
C J' = (J  - J' C  J')^{-1} J' =  (J'^{-1} J  -  C  J')^{-1} ,
\end{equation} 
or with $X=CJ'$ and $B= J'^{-1} J$ we arrive at the quadratic matrix equation
\begin{equation}
X^{2} -BX -1 =0.
\label{xeq}
\end{equation}

The solution of a quadratic matrix equation of the form $A X^{2} + B X + C$, in general, cannot be given
in a closed form unless $A=1$, $B$ commutes with $C$, and the square root of $B^{2} - 4C$ exists.
In this case the solution is given by
\begin{equation}
X_{\pm} = -B/2 \pm \sqrt{ B^{2} - 4C}/2.
\end{equation}

Obviously our Eq.\ (\ref{xeq}) meets this condition, and so the equation can be solved and $C$ can be calculated,
although the final formula is a little lengthy to present here. Nevertheless  a closed form expression has been obtained for the
tail. Starting with $C_{N'+1} \simeq C/N'$ the matrix continued fraction becomes convergent for the whole complex energy plane
and thus the method became amenable for calculating resonant states.

\subsection{Numerical illustration }

Here we adopt units such that $m =\hbar = e^{2} = 1$ and $c = 137.036$.
As a numerical illustration we consider the potential
\begin{equation}
V(r) = {92 }/{r} - 240 { \exp(-r)}/{r } +320 { \exp(-4 r)}/{r }. 
\end{equation}
This potential, for an  $l=0$ partial wave, has one bound state and a very narrow resonant state. We used the parameter $b=8$.  
The non-relativistic energies are
 $  -5.9293680 $  and  $ 15.6091791  -0.0000015 i  $, respectively, while the relativistic ones are 
 $  -5.9335096   $ and $ 15.5994090  -0.0000004  i$. We see that the method can calculate
resonant states in a very accurate way such  that it is able to pinpoint the fine relativistic effects.

\section{Confinement potential }

In this section we consider confinement potentials. We assume that the confinement potential is scalar and the vector potential is Coulomb-like
\begin{equation}
\quad U = \alpha_{1} r  + \alpha_{2} r^{2}  +  v_{0}^{(s)} \quad \text{and}\quad \quad V = Z/ r  +   v_{4}^{(s)} .
\end{equation}
For the short range part of the Hamiltonian we find
\begin{equation}
H_{FV0}^{(s)} =    (\tau_{3}+i\tau_{2}) v_{0}^{(s)}   +   v_{4}^{(s)},
\end{equation}
and for the long range part we obtain
\begin{equation}
H_{FV0}^{(l)} =  (\tau_{3}+i\tau_{2}) \left(  \frac{p^{2}}{2m} + \alpha_{1} r  + \alpha_{2} r^{2}    \right) 
+\tau_{3} mc^{2} + \frac{Z}{r} .
\label{hlconf}
\end{equation}
We have seen before that the representation of the constant $E$ and the $p^{2}$ are tridiagonal infinite 
matrices, but the confining $r$ term is penta-diagonal and the $r^{2}$ term is septa-diagonal. 
Consequently, the $\tau_{i}$ matrices of the FV0 equation
get superimposed on a septa-diagonal structure. As a result, we obtain an infinite tridiagonal matrix 
with $6 \times 6$ blocks and the procedure of Eqs.\  (\ref{G1}) and (\ref{cf}) for the Green's operator 
is applicable with $6\times 6$ blocks. We can also see that in the $N'\to \infty$ limit the $r$ term behaves like $N'^{2}$ and 
the $r^{2}$ term like $N'^{3}$.  Therefore,  
the confining terms dominate over the energy and kinetic energy terms. 
Hence, as $N'\to \infty$,  $C_{N'+1} \simeq 1/N'^{2}$ or $C_{N'+1} \simeq 1/N'^{3}$.
In either case,  the evaluation of the matrix continued fraction can be 
initiated with $C_{N'+1} = 0$.

\subsection{Numerical illustration }

We assume here that the Hamiltonian is in the form of Eq.\ (\ref{hlconf}) and there is no short range term. 
Table \ref{table2} shows the non-relativistic and relativistic results for the few lowest bound states of the system
with $Z=-1$ and $\alpha_{1}=1$, as well as $Z=-1$ and $\alpha_{2}=1/2$.
The former case is called the Cornell potential, and it is typically used in describing quarks confined in hadrons, while
the later one is a Coulomb plus harmonic oscillator potential. 
We can see in Table \ref{table2} that the method can pinpoint the fine relativistic effects.

\begin{table}[ht]
\caption{Non-relativistic and relativistic FV0 results for states with Coulomb ($Z=-1$) plus linear ($\alpha_{1}=1$) and 
quadratic ($\alpha_{2}=1/2$) confinement potentials.}
\label{table2}
\begin{tabular}{|c|c|c|c|c|}     
\hline
  Sch: $\alpha_{1}=1$ & FV0: $\alpha_{1}=1$& Sch: $\alpha_{2}=1/2$&  FV0: $\alpha_{2}=1/2$ \\
\hline
\hline
	0.57792135   &	0.57774937   &	0.17966848  &	 0.15989685	  \\
	2.45016289  &	2.44983403  &	2.50000000  &	 2.37624749	  \\
	3.75690569  &	3.75635589  &	4.63195241  &	 4.33778167	  \\
	4.85567124  &	4.85486537  &	6.71259573  &	 6.18557261	  \\
	5.83602989 &	5.83494151  &	8.76951960  &	 7.93366119 	  \\
	6.73662100 &	6.73522824  &   10.8129243 &	 9.56883321	  \\
  \hline
\end{tabular}\label{table:accuracy_check}
\end{table}

\section{Summary and conclusions}

There are a great deal of methods addressing problems in non-relativistic quantum mechanics, but
much fewer methods are available for relativistic systems. In this work we addressed relativistic problems through the
Feshbach-Villars formalism. In this formalism we dealt with Schr\"odinger-like Hamilton eigenvalue problems.
This formalism features a multicomponent wave function  with kinetic energy
coupling the components. In this approach we cast the FV0 equations into an integral equation, which is represented in
a discrete basis. The corresponding Green's operator is calculated for two important long-range potentials,  the Coulomb
and confining potentials. The multicomponent character with the kinetic energy and long range terms as asymptotic
couplings were successfully treated with the help of the matrix continued fractions. 
By calculating the tail we managed  to extend the methods for complex energies as well.

\bibliography{inversion}

\end{document}